\documentclass[english]{article}
\usepackage[T1]{fontenc}
\usepackage{float}
\usepackage{amsmath}
\usepackage{graphicx}
\usepackage{esint}
\PassOptionsToPackage{normalem}{ulem}
\usepackage{ulem}

\makeatletter
\usepackage{enumitem}
\setenumerate{label=(\roman*)}
\usepackage{cite}

\usepackage{babel}

\usepackage{babel}

\makeatother

\usepackage{babel}
\begin{document}

\title{Fluctuation induced forces in the presence of mobile carrier drift}

\author{Boris Shapiro\\
Department of Physics, Technion-Israel Institute of Technology,\\
Haifa 32000, Israel}
\maketitle
\begin{abstract}
A small polarizable object (an atom, molecule or nanoparticle), placed above a medium with flowing dc current in it, is considered. It is shown that the dc current can have a strong effect on the force exerted on the particle. The Casimir-Lifshitz force, well studied in the absence of current, gets modified due to drifting mobile carriers in the medium. Furthermore, a force in the lateral direction appears. This  force is a non-monotonic  function of the drift velocity and its maximal value is comparable with the Casimir-Lifshitz force. If the temperatures of the medium and the particle are different, this lateral force can be directed along the current (drag) or in the opposite direction (anti-drag).

\end{abstract}
\newpage{}
\section{Introduction}

All bodies are surrounded by a fluctuating electromagnetic field,
due to the random motion of charges inside a body. When a second
body is placed in the vicinity of the first one,
fluctuation-induced (Casimir-Lifshitz) forces appear between the
bodies.
These forces are of great relevance in chemistry, nanotechnology
and biology \cite{bor, par}. Much of the recent work on the
fluctuation-induced forces, as well as on the related phenomena of
near field heat transfer and the noncontact friction, deals with
systems out of equilibrium (for some reviews see \cite{bim,volo,
intra,dek,kya,ded,jou}). One should distinguish among several
out-of-equilibrium situations :
\begin{enumerate}
\item Different parts of the system have different temperatures
but there is no relative motion between those parts (a hot body
embedded into the cold environment is the simplest example of such
situation \cite{jou,lev,ryt}). Under such conditions the
Casimir-Lifshitz forces will be modified, as compared to their
equilibrium value \cite{bim,volo,
intra,dek,kya,ded,ant,bim',krug,lu,bart,dru}.\label{enu:Different-parts-of-1}
\item Different parts of the system are in relative motion. For
instance, two macroscopic plates, separated by a vacuum gap, move
one on top of the other. Another example is an atom (or a
nanoparticle) moving above a macroscopic plate. Relative motion
between bodies affects the Casimir-Lifshitz forces and, in
particular, leads to dissipation and noncontact friction. This
kind of problems was considered by many authors (\cite{bim,volo,
intra,dek,kya,ded,intra',teo,fer,tom,pen,vol} and references
therein), with rather controversial results (see Ref \cite{kya}
for various contradictions and inconsistencies in the
literature).\label{enu:Different-parts-of} \item There is no
relative motion between parts of the system but some of the parts
are subjected to a dc electric current \cite{bs,bs',vol',vol'',dean} . The simplest example is to
consider a semiconducting plate, with a dc current flowing in it,
and ask how this current affects fluctuations of the
electromagnetic field inside and outside the plate. This problem
has been considered in \cite{bs,bs'}. In the present paper we
further elaborate on electromagnetic field fluctuations in the
presence of carrier drift and, in particular, calculate the
fluctuation force acting on an atom, or a nanoparticle placed
above a sample with a dc current in it.\label{enu:There-is-no}
\end{enumerate}
Let us stress that setups \ref{enu:Different-parts-of} and
\ref{enu:There-is-no} are quite different- a fact not sufficiently
appreciated in the literature.
For one thing,
the dc current in the sample produces a stationary (time
independent) magnetic field which affects the atomic spectrum and,
if inhomogeneous, exerts a force on the atom as a whole. More
importantly, the fluctuation-induced forces in the two setups are
not the same. The point is that in setup \ref{enu:There-is-no} the
mobile carriers are in motion (in the laboratory frame) while the
lattice is fixed. Therefore the spontaneous fluctuations
originating in the sub-system of the mobile carriers will be
Doppler shifted with respect to those residing in the lattice.
Moreover, in the presence of drift it is generally not even
possible to assign a definite temperature to the mobile carriers,
which makes the existing theory of the fluctuation-induced forces
inapplicable. The purpose of this work is to study the effect of
carrier drift on the fluctuation forces exerted on a small
polarizable object (an atom, molecule or a nanoparticle).

The organization of the paper is as follows: In Section II  we define the model and discuss how the fluctuational
electrodynamics (Rytov's theory) should be modified in the presence of mobile carrier drift. Section III is devoted to the properties of the fluctuating field, outside the sample with drifting carriers. In Section IV a particle is introduced, above the surface of a medium with drifting carriers, and the forces acting on the particle (both in the lateral and in the normal direction) are calculated. Various specific examples are presented in Section
V and the conclusions are summarized in Section VI.




\section{Fluctuational Electrodynamics in the Presence of Carrier Drift\label{sec:Flactuational-Electrodynamics-in}}

We consider a conducting medium, e.g., a semiconductor, containing
mobile carriers with charge $e$, effective mass $m$ and equilibrium
concentration $n_{0}$. When a dc voltage is applied to the sample,
the carriers acquire some drift velocity $\vec{v}_{0}$ , so that there
is a steady state dc current $\vec{j}_{0}=en_{0}\vec{v}_{0}$. On
top of this stationary drift there are fluctuations of the carrier
and current density which cause fluctuations of the electric
field. We designate the fluctuating part of these quantities as $\vec{n}\left(\vec{r},t\right)$
, $\vec{j}\left(\vec{r},t\right)$ and $\vec{E}\left(\vec{r},t\right)$,
respectively. Thus, for instance, the total current density is $\vec{j}_{0}+\vec{j}\left(\vec{r},t\right)$
. The fluctuating part $\vec{E}\left(\vec{r},t\right)$ of the electric
field is of particular interest because, unlike $n$ and $\vec{j}$,
it exists also outside the sample and exerts forces on nearby objects.
It should be emphasized that $\vec{j}\left(\vec{r},t\right)$ accounts
only for the motion of the mobile carriers. In addition, there are
fluctuating polarization currents due to the lattice. We briefly recapitulate
the main equations of the theory, following with some modifications
Ref \cite{bs'}.

The relation between $\vec{j}$ and $\vec{E}$, in the frequency-wavevector domain is
\begin{equation}
j_{\alpha}\left(\omega,\vec{k}\right)=\sigma_{\alpha\beta}\left(\omega,\vec{k}\right)E_
{\beta}\left(\omega,\vec{k}\right),\label{ohm}
\end{equation}
where summation over $\beta$ is implied. The conductivity tensor
$\sigma_{\alpha\beta}$ is defined with respect to the
non-equilibrium steady state, i.e., it connects quantities
fluctuating on top of the stationary current flow. That is why,
even for an intrinsically isotropic medium, $\sigma_{\alpha\beta}$
is a tensor depending not only on $\omega$ but also on $\vec{k}$.
The dependence on $\vec{k}$ occurs because the fluctuations are
carried away by the flow, thus producing a non-local response
(spatial dispersion). Adding the conduction current,
 Eq (\ref{ohm}), to the fluctuating polarization current of the lattice
yields the fluctuating displacement
\begin{equation}
\begin{array}{c}
D_{\alpha}\left(\omega,\vec{k}\right)=\epsilon_{L}\left(\omega\right)E_{\alpha}
\left(\omega,\vec{k}\right)+i\frac{4\pi}{\omega}\sigma_{\alpha\beta}\left(\omega,\vec{k}\right)E_{\beta}\left(\omega,\vec{k}\right)\\
\equiv\epsilon_{\alpha\beta}\left(\omega,\vec{k}\right)E_{\beta}\left(\omega,\vec{k}\right),
\end{array}\label{displacement}
\end{equation}
where $\epsilon_{L}$ is the lattice dielectric function which can
depend on $\omega$ but not on $\vec{k}$. Eq (\ref{displacement})
defines the dielectric tensor
$\epsilon_{\alpha\beta}\left(\omega,\vec{k}\right)$ which controls
the dynamics of electrical fluctuations in the medium. The form of
$\epsilon_{\alpha\beta}\left(\omega,\vec{k}\right)$ depends on the
specific system or model. We assume here Drude model, with drift,
which is a special case of the more general hydrodynamic model
(see Eq(11) of \cite{bs'} with the thermal pressure term
neglected) :
\begin{equation}
\epsilon_{\alpha\beta}\left(\omega,\vec{k}\right)=\left(\epsilon{}_{L}^{\prime}+i\epsilon^{\prime\prime}{}_{L}\right)
\delta_{\alpha\beta}-\frac{\omega_{p}^{2}}{\omega\left(\omega-\vec{k}\cdot\vec{v}_{0}+i\nu\right)}\left(\delta_{\alpha\beta}+
\frac{v_{0\alpha}k_{\beta}}{\omega-\vec{k}\cdot\vec{v}_{0}}\right),\label{epsilon}
\end{equation}
where $\nu$ is the collision frequency of the mobile carriers, $\omega_{p}^{2}=4\pi e^{2}n_{0}/m$
, and $\epsilon_{L}$ has been separated into the real and imaginary
parts.

In our dealing with fluctuations we use Rytov's method
in which random Langevin sources are introduced into the Maxwell equations,
similarly to what is done in the theory of Brownian motion. These
random sources play the role of  "external" currents and charges
in the Maxwell equations and, if their correlation functions are known,
one can compute the correlation function for various components of
the electromagnetic field. We shall be interested in fluctuational
phenomena close to the surface of the sample and neglect the retardation
effects.
In
this limit the electromagnetic field is rotationless, $\vec{E}\left(\vec{r},t\right)=-\nabla\Phi\left(\vec{r},t\right)$, and Rytov's fluctuational electrodynamics reduces to the Poisson
equation supplemented by the Langevin sources. In the bulk of the
sample this equation is
\begin{equation}
k^{2}\epsilon\left(\omega,\vec{k}\right)\Phi\left(\omega,\vec{k}\right)=
4\pi\rho_{r}\left(\omega,\vec{k}\right),\label{pois}
\end{equation}
where $\Phi\left(\omega,\vec{k}\right)$ , $\rho_{r}\left(\omega,\vec{k}\right)$
are the Fourier transforms \cite{foot1} of the potential $\Phi\left(\vec{r},t\right)$
and of the random Langevin sources $\rho_{r}\left(\vec{r},t\right)$,
and
\begin{equation}
\epsilon\left(\omega,\vec{k}\right)=\frac{k_{\alpha}k_{\beta}}{k^{2}}\epsilon_{\alpha\beta}\left(\omega,\vec{k}\right)=
\epsilon_{L}^{\prime}\left(\omega\right)+i\epsilon_{L}^{\prime\prime}\left(\omega\right)-\frac{\omega_{p}^{2}}
{\left(\omega-\vec{k}\cdot\vec{v}_{0}+i
\nu\right)\left(\omega-\vec{k}\cdot\vec{v}_{0}\right)}\label{epsilons}
\end{equation}
Thus, the tensorial dielectric function in Eq (\ref{epsilon}) reduces to a scalar. (If the retardation effect were taken into account, then the full dielectric function, Eq (\ref{epsilon}), would come into play.)
This expression has a simple interpretation. The dielectric function $\epsilon\left(\omega,\vec{k}\right)$ relates the displacement and the field in a longitudinal wave. If the wave propagates in the direction of flow, $\vec{k}\parallel \vec{v}_{0}$, then there is a Doppler shift of the wave frequency. There is no such shift if the propagation direction is perpendicular to $\vec{v}_{0}$. Note that only the plasma component of $\epsilon\left(\omega,\vec{k}\right)$ undergoes the Doppler shift, while the lattice component remains the same as in equilibrium.

For a system at equilibrium ($\vec{j}_{0}=0$) the correlation function
of the random sources is determined by the fluctuation-dissipation
theorem \cite{lev,ryt} :
\begin{equation}
\begin{array}{c}
\left\langle \rho_{r}\left(\omega,\vec{k}\right)\rho_{r}^{*}\left(\omega^{\prime},\vec{k^{\prime}}\right)\right\rangle =
2\pi\delta\left(\omega-\omega^{\prime}\right)\left\langle \rho_{r}\left(\vec{k}\right)\rho_{r}^{*}\left(\vec{k^{\prime}}\right)\right\rangle _{\omega}\\
=\left(2\pi\right)^{4}\delta\left(\omega-\omega^{\prime}\right)\delta\left(\vec{k}-\vec{k^{\prime}}\right)\left\langle
\rho_{r}\rho_{r}^{*}\right\rangle _{\omega\vec{k}}
\end{array}
\label{corr}
\end{equation}
with
\begin{equation}
\left\langle \rho_{r}\rho_{r}^{*}\right\rangle _{\omega\vec{k}}=\frac{\hbar k^{2}}{4\pi}\epsilon^{\prime\prime}\left(\omega\right)
\coth\frac{\hbar\omega}{2T} ,\label{fdt}
\end{equation}
where $\left\langle \dotsb\right\rangle $ denotes thermal and
quantum average, $T$ is the temperature of the system and
$\epsilon^{\prime\prime}\left(\omega\right)$ is the imaginary part
of its dielectric function [Eq (\ref{epsilons}) with
($\vec{v}_0=0$)].   Eq (\ref{corr}) defines the spectral densities
$\left\langle
\rho_{r}\left(\vec{k}\right)\rho_{r}^{*}\left(\vec{k^{\prime}}\right)\right\rangle
_{\omega}$ , $\left\langle \rho_{r}\rho_{r}^{*}\right\rangle
_{\omega\vec{k}}$ , and Eq (\ref{fdt}) contains the essence of the
fluctuation-dissipation
theorem. 
Strictly speaking, $\rho_{r}$, $\vec{j}\left(\vec{r},t\right)$,
$\vec{E}\left(\vec{r},t\right)$, etc., should be understood as
quantum-mechanical operators and various correlation functions
should be properly antisymmetrized. These changes, however, would
be only \textquotedbl{}cosmetic\textquotedbl{} and would not
affect the final results. The point is that in the RHS of Eq
(\ref{fdt}) the correct quantum mechanical spectral density is
given. With this ceavet, Rytov's theory becomes essentially
classical \cite{LL}.

Since our system is out of equilibrium ($\vec{j}_{0}\ne0$), there
is no general prescription for writing down the correlator of the
random sources $\rho_{r}\left(\omega,\vec{k}\right)$. However, under
some conditions, it is possible to do so.
As far as the lattice is concerned, the use of the fluctuation-dissipation theorem is justified because the lattice, even in the presence of an electric current, is usually close to equilibrium, i.e., the phonon distribution is close to Bose-Einstein. More precisely, the lattice is in internal equilibrium with some temperature $T_L$, generally different from the environment temperature. Such internal equilibrium is a sufficient condition for applying the fluctuation-dissipation relation, as is indeed done in all the work where Casimir-Lifshitz forces or heat flow between bodies at different temperatures are considered. Thus, for the random sources originating in the lattice one can use the equilibrium theory, Eqs (\ref{corr}), (\ref{fdt}),  with
$\epsilon^{\prime\prime}\left(\omega\right)$
 replaced by $\epsilon_{L}^{\prime\prime}\left(\omega\right)$ \cite{bs',foot2,foot3}

\begin{equation}
\left\langle \rho_{r}\left(\vec{k}\right)\rho_{r}^{*}\left(\vec{k^{\prime}}\right)\right\rangle _{\omega}^{L}=\frac{\hbar k^{2}}
{4\pi}\left(2\pi\right)^{3}\delta\left(\vec{k}-\vec{k^{\prime}}\right)\epsilon_{L}^{\prime\prime}\left(\omega\right)\coth\frac{\hbar\omega}{2T_{L}}
\label{corrA}
\end{equation}

Similarly, in order to apply the fluctuation-dissipation theorem to the drifting plasma one should require that the plasma is in internal equilibrium, in its own frame of reference.
 This
happens, for instance, at low drift velocities, when the electron
distribution is close to Fermi-Dirac (or Bolzmann), with the
temperature of the lattice. The more interesting example is the
case of large drift velocities when, due to strong mutual
interactions, the electronic system undergoes rapid internal
thermalization, with a temperature higher than that of the lattice
("hot electrons"). We assume that the condition of internal equilibrium with some temperature $T_{el}$ is satisfied. 
In this
case the random sources residing  in the plasma are controlled by the imaginary part
of the last term in Eq (\ref{epsilons}).
Denoting this term by $\epsilon_{el}(\omega_{-})$, where $\omega_{-} = \omega-\vec{k}\cdot\vec{v}_{0}$, we
can write the fluctuation-dissipation theorem as
\begin{equation}
\left\langle
\rho_{r}\left(\vec{k}\right)\rho_{r}^{*}\left(\vec{k^{\prime}}\right)\right\rangle
_{\omega}^{el} =\frac{\hbar
k^{2}}{4\pi}\left(2\pi\right)^{3}\delta\left(\vec{k}-\vec{k^{\prime}}\right)\epsilon_{el}^{\prime\prime}\left(\omega_{-}\right)
\coth\frac{\hbar\omega_{-}}{2T_{el}}\label{corrB}
\end{equation}

The important difference between the Eqs (\ref{corrA}) and
(\ref{corrB}), besides the trivial replacement of
$\epsilon_{L}^{\prime\prime}$, $T_L$ by
$\epsilon_{el}^{\prime\prime}$, $T_{el}$, is that the frequency
$\omega_{-}$ appears in Eq (\ref{corrB}), i. e., the spontaneous
creation of the fluctuations is now affected by the drift: the
frequency of the fluctuations, as measured in the laboratory
frame, is Doppler shifted. Let us note in this context that, while the electronic part of the dielectric function depends on the Doppler shifted frequency  $\omega_{-}$, the lattice part depends  on the "bare" frequency $\omega$.
Therefore  the problem of fluctuations in the presence of
carrier drift is not equivalent to that for a moving sample. Only
if one makes the additional assumption that $\epsilon_{L}
= const$
do the two problems become equivalent (provided that the drifting electrons are in an internal equilibrium, which is in itself  a rather strong assumption).

The two contributions to the spontaneous random sources [Eqs (\ref{corrA}), (\ref{corrB})] are, of course, uncorrelated since they originate in two different subsystems- the lattice and the electron plasma. These equations, together with the Poisson equation and the expression for the dielectric function, Eq (\ref{epsilons}), allows us to treat fluctuations of various quantities, both inside and outside the medium with current, in the quasistatic limit (to include the retardation effects one has to replace the Poisson equation by the full set of Maxwell equations). Throughout the paper we discuss separately two limiting models, when either the lattice or the plasma make the dominant contribution to the fluctuations. While in principle it would be possible to consider the general situation, when both components make a comparable contribution, this would make the already combersome equations even more complicated and would only blur the basically simple physical picture.

Later, when considering the phenomena near a planar surface of the
medium, we shall need Eq (\ref{pois}) in a somewhat different form.
Assuming that the velocity vector $\vec{v}_{0}$ is in the $\left(x,y\right)$
- plane , i.e., $\epsilon\left(\omega,\vec{k}\right)$ does not depend
on $k_{z}$ , we can transform Eq (\ref{pois}) back to space, in the
$z$ direction, obtaining
\begin{equation}
\epsilon\left(\omega,\vec{q}\right)\left(-\frac{\partial^{2}}{\partial z^{2}}+q^{2}\right)\Phi\left(\omega,\vec{q},z\right)
=4\pi\rho_{r}\left(\omega,\vec{q},z\right),\label{pois'}
\end{equation}
where $\vec{q}=\left(k_{x},{k}_{y}\right)$ denotes the transverse
(in-plane) wave vector and $\Phi\left(\omega,\vec{q},z\right)$ is
the Fourier transform of $\Phi\left(x,y,z,t\right)$ with respect
to time and the $x,y$ - coordinates (the same for $\rho_{r}$).
This $(\omega,\vec{q},z)$-representation is convenient for handling the planar geometry. The
spectral densities of the random sources, Eqs (\ref{fdt},\ref{corrA},\ref{corrB}), should be also transformed to
the $(\omega,\vec{q},z)$-representation. For instance, Eq (\ref{corrA}) becomes
\begin{multline}
\left\langle \rho_{r}\left(\vec{q},z\right)\rho_{r}^{*}\left(\vec{q^{\prime}},z^{\prime}\right)\right\rangle _{\omega}^{L}=\frac{\hbar}{4\pi}\left(2\pi\right)^{2}\delta\left(\vec{q}-\vec{q^{\prime}}\right)\left[q^{2}\delta\left(z-z^{\prime}\right)+\frac{\partial^{2}}{\partial z\partial z^{\prime}}\delta\left(z-z^{\prime}\right)\right]\\
\times\epsilon_{L}^{\prime\prime}\left(\omega\right)\coth\frac{\hbar\omega}{2T_{L}},\label{corrA'}
\end{multline}
and similarly for the other spectral densities.


\section{Fluctuations of the Electric Potential Near the Surface}

We consider a medium occupying half space $\left(z<0\right)$ while
the other half $\left(z>0\right)$ is vacuum. The random charge
sources $\rho_{r}\left(\vec{r},t\right)$ produce evanescent
electric fields near the surface (in addition to the radiation
which we do not consider within our quasi-stationary, non-retarded
approximation) and we are interested in various correlation
functions for the potential and field. We shall consider three
different setups, see Fig 1. Although case (a) has been studied
long ago \cite{jou,lev,ryt,car,hen} and case (b) is simply related
to (a), we discuss briefly also these two cases. The correlation
functions obtained in this section will serve as building blocks
in calculation of the fluctuation-induced forces in the next
section.

\begin{figure}[H]
\includegraphics[scale=0.5]{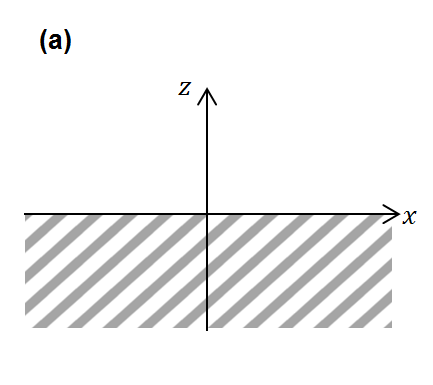}\includegraphics[scale=0.5]{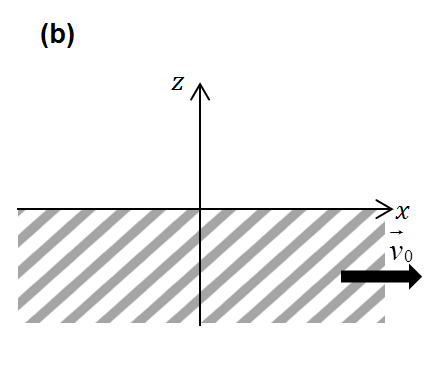}
\includegraphics[scale=0.5]{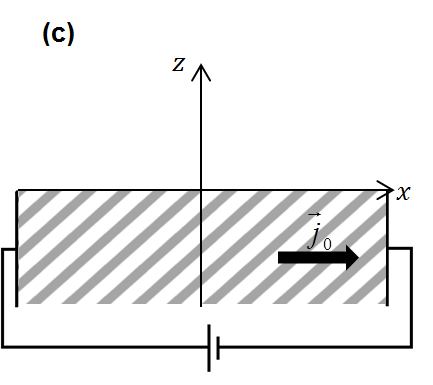}\caption{The three setups: (a) The medium is at rest (in the laboratory frame),
at equilibrium. (b) The medium moves with a constant velocity $\vec{v}_{0}$
, along the $x$- axis. (c) The medium is at rest but a dc current
with density $\vec{j}_{0}=en_{0}\vec{v}_{0}$ is flowing in the medium. }
\end{figure}

Case (a) : The equilibrium  dielectric function is (Eq
 (\ref{epsilons}) with $\vec{v}_{0}=0$ )
\begin{equation}
\epsilon\left(\omega\right)=\epsilon_{L}^{\prime}\left(\omega\right)+i\epsilon_{L}^{\prime\prime}\left(\omega\right)
-\frac{\omega_{p}^2}{\left(\omega+i\nu\right)\omega},\label{eps}
\end{equation}
and we have to solve Eq (\ref{pois'}) with this expression for $\epsilon\left(\omega,\vec{q}\right)$.
If one defines the Green's function
\begin{equation}
\epsilon\left(\omega\right)\left(-\frac{\partial^{2}}{\partial z^{2}}+q^{2}\right)g\left(z,z_1,\vec{q},\omega\right)=4\pi\delta
\left(z-z_1\right)\label{green}
\end{equation}
then the formal solution of Eq (\ref{pois'}) is
\begin{equation}
\Phi\left(\omega,\vec{q},z\right)=\int_{-\infty}^{0}dz_1g\left(z,z_1,\vec{q},\omega\right)
\rho_{r}\left(\omega,\vec{q},z_1\right).\label{Phi}
\end{equation}
The solution of Eq (\ref{green}), with the source inside the medium
$\left(z_1<0\right)$, the observation point outside $\left(z>0\right)$
and the standard boundary conditions for $\Phi$ and its normal derivative
at $z=0$, is
\begin{equation}
g\left(z,z_1,\vec{q},\omega\right)=
\frac{4\pi}{q}\frac{1}{\epsilon\left(\omega\right)+1}e^{-q\left(z-z_1\right)}\label{green'}
\end{equation}

Since in equilibrium
the lattice temperature $T_{L}$ is the same as the electron
temperature, we can leave $T_{L}$ to denote the equilibrium temperature of the
sample. Then, using (\ref{Phi}),(\ref{green'}) and (\ref{corrA'}) (with $\epsilon^{\prime\prime}\left(\omega\right)$
insted of $\epsilon_{L}^{\prime\prime}\left(\omega\right)$ ) one
obtains after some algebra:
\begin{equation}
\left\langle \Phi\left(\omega,\vec{q},z\right)\Phi^{*}\left(\omega^{\prime},\vec{q^{\prime}},z^{\prime}\right)\right\rangle =
\left(2\pi\right)^{3}\delta\left(\omega-\omega^{\prime}\right)\delta\left(\vec{q}-\vec{q^{\prime}}\right)\left\langle \Phi\left(z\right)
\Phi^{*}\left(z^{\prime}\right)\right\rangle _{\omega\vec{q}}\label{Phycorra}
\end{equation}
with
\begin{equation}
\left\langle \Phi\left(z\right)\Phi^{*}\left(z^{\prime}\right)\right\rangle _{\omega\vec{q}}=4\pi\hbar\frac{\epsilon^{\prime\prime}\left(\omega\right)}{\left|
\epsilon\left(\omega\right)+1\right|^{2}}\coth\left(\frac{\hbar\omega}{2T_{L}}\right)\frac{1}{q}e^{-q\left(z+z^
{\prime}\right)}\label{Phycorrb}
\end{equation}
Expression (\ref{Phycorrb}) factorizes into the $\omega$-dependent
and $q$ - dependent parts. The Fourier transform from $\vec{q}=\left(k_{x},k_{y}\right)$
to $\vec{\rho}=\left(x,y\right)$ immediately yields
\begin{equation}
\left\langle
\Phi\left(x,y,z\right)\Phi^{*}\left(x^{\prime},y^{\prime},z^{\prime}\right)\right\rangle
_{\omega}=2\hbar\frac{\epsilon^{\prime\prime}\left(\omega\right)}{\left|\epsilon\left(\omega\right)+1\right|^{2}}
\coth\left(\frac{\hbar\omega}{2T_{L}}\right)\frac{1}{\sqrt{\left(x-x^{\prime}\right)^{2}+\left(y-y^{\prime}
\right)^{2}+\left(z+z^{\prime}\right)^{2}}}.\label{Phycorrc}
\end{equation}
The correlation function $\left\langle \Phi\left(x,y,z,\omega\right)\Phi^{*}\left(x^{\prime},y^{\prime},z^{\prime},\omega^{\prime}\right)\right\rangle $
is obtained from
(\ref{Phycorrc})
by multiplying it by the factor $2\pi\delta\left(\omega-\omega^{\prime}\right)$.
This is the general rule, for any pair of fluctuating variables, and
it follows from the stationary character of the fluctuations.

Correlation
functions for various components of the electric field can be obtained
from Eq
(\ref{Phycorrc})
 by differentiation. Some examples can be found
in the above cited literature. For instance, differentiating
 Eq (\ref{Phycorrc})
with respect to $x$ and $x^{\prime}$ and setting at the end $x^{\prime}=x$,
one obtains
\begin{equation}
\left\langle E_{x}^{2}\right\rangle _{\omega}=\frac{\hbar}{4z^{3}}\frac{\epsilon^{\prime\prime}\left(\omega\right)}{\left|
\epsilon\left(\omega\right)+1\right|^{2}}\coth\left(\frac{\hbar\omega}{2T_{L}}\right),\label{E}
\end{equation}
i.e., when the surface is approached, the energy density increases
as $\frac{1}{z^{3}}$ - the well known rule. This rule breaks down,
of course, for sufficiently small $z$, either because of spatial
dispersion effects or simply because the macroscopic theory becomes
inapplicable at atomic distances.

Case (b): The medium is now moving, in the laboratory frame, with
velocity $v_{0}$ in the $x$ direction. In the frame moving with
the medium all the relations derived above for case (a) remain of
course valid, if the coordinates and frequency refer to that
frame. It is immediate to translate the results to the laboratory
frame. If we denote by $F(x-x', y-y', z, z', t-t')$ some
correlation function at equilibrium (i.e., in the rest frame of
the sample), then the correlation function in the laboratory frame
is simply $F_{lab}(x-x', y-y', z, z', t-t') = F(x- v_0t -x' +
v_0t', y-y', z, z', t-t')$. This relation holds in the
non-relativistic limit, $v_{0}\ll c$ , assumed in the present
work, and it implies that the Fourier transform
$F_{lab}\left(\omega, k_x, k_y, z, z' \right)$ is obtained from
$F\left(\omega, k_x, k_y, z, z' \right)$ by replacing $\omega$
with $\omega- k_xv_{0}\equiv\omega_{-}$. For instance the spectral
density $<\Phi(z)\Phi^{*}(z^{\prime}> _{\omega\vec{q}}$ for a
moving sample, as viewed from the laboratory frame, is given by
the same expression as in the right-hand-side of
Eq.(\ref{Phycorrb}) but with $\omega_{-}$ instead of
$\omega$. Note, though, that this replacement results in a
complicated, non-factorizable function of $k_x, k_y$ and $\omega$,
and no simple expression in real space, comparable to Eq
(\ref{Phycorrc}), can be obtained. We will not elaborate on this
case further but move on to

Case (c): Here the sample is at rest but the electron subsystem
moves with respect to the lattice with velocity $\vec{v}_{0}$,
producing a dc current density $\vec{j}_{0}=en_{0}\vec{v}_{0}$. We
do not specify the model for the lattice but just describe it by
the lattice constant
$\epsilon_{L}=\epsilon_{L}^{\prime}\left(\omega\right)+i\epsilon_{L}^{\prime\prime}\left(\omega\right)$.
The subsystem of the mobile carriers is described by the Drude
model with drift, Eq (\ref{epsilons}).

In the limit of small collision frequency $\nu$ (the collisionless
plasma model) the Langevin sources originating in the lattice
dominate over those in the electronic subsystem. Taking the latter
as ``noiseless'' and assuming the lattice in equilibrium, we can
study the fluctuations using Eqs. (\ref{pois'}),(\ref{corrA'})
with
\begin{equation}
\epsilon\left(\omega,\vec{k}\right)=\epsilon_{L}^{\prime}(\omega)+i\epsilon_{L}^{\prime\prime}(\omega)-
\frac{\omega_{p}^{2}}{\omega_{-}^{2}}\equiv
\epsilon_{1}\left(\omega,k_{x}\right).\label{epsA}
\end{equation}
 This model has been
considered in \cite{bs'}. An unnecessary approximation was
introduced there at an early stage of the calculation. Here we
present a somewhat different approach.

In fact, in $\left(\vec{q},z\right)$ - representation (i.e.,
Fourier transform in the $\left(x,y\right)$ - plane but not in the
$z$ - direction) the calculation is straightforward and almost
identical to case (a).
The only difference is that the dynamics of the fluctuations is
now controlled by the $(k_x)$-dependent dielectric function in Eq
(\ref{epsA}), so that instead of (\ref{Phycorrb}) we  have
\begin{equation}
\left\langle
\Phi\left(z\right)\Phi^{*}\left(z^{\prime}\right)\right\rangle
_{\omega\vec{q}}=4\pi\hbar\frac{\epsilon_{L}^{\prime\prime}\left
(\omega\right)}{\left|\epsilon_{1}\left(\omega,k_{x}
\right)+1\right|^{2}}\coth\left(\frac{\hbar\omega}{2T_{L}}\right)\frac{1}{q}e^{-q\left(z+z^{\prime}\right)}\label{c}
\end{equation}
and the desired spectral density is
\begin{equation}
\left\langle
\Phi\left(x,y,z\right)\Phi^{*}\left(x^{\prime},y^{\prime},z^{\prime}\right)\right\rangle
_{\omega}=\int\int\frac{dk_{x}dk_{y}}{\left(2\pi\right)^{2}}e^{ik_{x}\left(x-x^{\prime}\right)+ik_{y}\left(y-y^{\prime}\right)}\left\langle
\Phi\left(z\right)\Phi^{*}\left(z^{\prime}\right)\right\rangle
_{\omega\vec{q}}.\label{c'}
\end{equation}

Because of the $(k_x)$ dependence of $\epsilon_1$
the integrand in (\ref{c'}) does not factorize, as it did in case
(a), and no simple analytical expression
can be obtained. For small drift velocities one can expand
$\epsilon_{1}\left(\omega,k_{x}\right)$
in powers of $\left(\frac{k_{x}v_0}{\omega}\right)$. The first power
does not contribute to (\ref{c'}) due to symmetry. The second
power contributes, e.g., to the quantity $\left\langle
E_{x}^{2}\right\rangle _{\omega}$ in Eq (\ref{E}), a term
proportional to $\left(v_{0}^{2}/z^{5}\right).$ This follows from
a simple power counting: an extra factor $k_{x}^{2}$ in the integrand contributes an extra term $\left(1/z^{2}\right)$ upon
integration over $k_{x}$.

It is worthwhile to mention an interesting qualitative effect due
to the drift. In equilibrium the spectral density, Eq (\ref{E}),
has a sharp maximum at the frequency of the surface plasmon
$\omega_{sp}=\omega_{p}/\sqrt{\epsilon_{L}^{\prime}+1}$, when the
factor $\left|\epsilon\left(\omega\right)+1\right|$ becomes close
to zero \cite{jou}. In the presence of drift we have
$\epsilon_{1}\left(\omega,k_{x}\right)$ instead of
$\epsilon\left(\omega\right)$, i.e., surface plasmons acquire
dispersion and, upon integration over $k_{x}$, the peak in
$\left\langle E_{x}^{2}\right\rangle _{\omega}$ gets broadened.

Let us write down a useful spectral function which will be needed
later:
\begin{multline}
\left\langle \vec{E}\left(x,y,z\right)\cdot\vec{E}^{*}\left(x^{\prime},y^{\prime},z^{\prime}\right)\right\rangle _{\omega}=8\pi\hbar\epsilon_{L}^{\prime\prime}\left(\omega\right)\coth\left(\frac{\hbar\omega}{2T_{L}}\right)\int\int\frac{dk_{x}dk_{y}}{\left(2\pi\right)^{2}}\\
\times
e^{ik_{x}\left(x-x^{\prime}\right)+ik_{y}\left(y-y^{\prime}\right)-q\left(z+z^{\prime}\right)}\frac{q}{\left|
\epsilon_{1}\left(\omega,k_{x}\right)+1\right|^{2}}\label{E'}
\end{multline}
 This result is obtained from Eq (\ref{c'}) {[}with Eq (\ref{c})
inserted{]} by differentiating with respect to the pairs of
variables
$\left(x,x^{\prime}\right),\left(y,y^{\prime}\right),\left(z,z^{\prime}\right)$
and adding the corresponding expressions.

This concludes our discussion of the case
when the lattice is the dominant source of noise. In the opposite
limit the
spontaneous random sources  occur predominantly in the electron plasma.
The appropriate dielectric function now is
\begin{equation}
\epsilon_{2}\left(\omega,k_{x}\right)=\epsilon_{L}^{\prime}(\omega)-\frac{\omega_{p}^{2}}{\omega_{-}\left(\omega_{-}+i\nu\right)}
\label{epsB}
\end{equation}
and the appropriate spectral density for the spontaneous random
sources is given in Eq. (\ref{corrB}) so that instead of Eq (\ref{c}) we have
\begin{equation}
\left\langle
\Phi\left(z\right)\Phi^{*}\left(z^{\prime}\right)\right\rangle
_{\omega\vec{q}}=4\pi\hbar\frac{\epsilon_{2}^{\prime\prime}\left
(\omega, k_x\right)}{\left|\epsilon_{2}\left(\omega,k_{x}
\right)+1\right|^{2}}\coth\left(\frac{\hbar\omega_{-}}{2T_{el}}\right)\frac{1}{q}e^{-q\left(z+z^{\prime}\right)}.\label{c"}
\end{equation}
This equation, unlike Eq (\ref{c}), contains $k_x$ not only in the dielectric function but also in the argument of the coth. Therefore the small velocity expansion has now a different structure: The first non-vanishing term is still quadratic in the parameter $\left(\frac{k_{x}v_0}{\omega}\right)$ but now one power can come from  the coth-function. Thus, the result will contain first derivative of the dielectric function, in addition to a term with the second derivative.



\section{Fluctuation - Induced Forces }

Consider an electric dipole, with dipole moment $\vec{p}$,
subjected to a space and time-dependent electromagnetic field
$\vec{E}\left(\vec{r},t\right), \vec{B}\left(\vec{r},t\right)$
. The size of the dipole is assumed to be much smaller than the
characteristic wave length of the field (a ``point dipole''). The
dipole can rotate or vibrate but it does not move as a whole,
i.e., it can be assigned a fixed position $\vec{r}_{0}$ and an
arbitrary time dependence $\vec{p}\left(t\right)$. Under such
conditions the dipole experiences an electric force
$\left(\vec{p}\cdot\nabla\right)\vec{E}\left(\vec{r},t\right)$
($\vec{r}$ is set equal to $\vec{r}_{0}$ after differentiation)
and the Lorenz magnetic force
$\left(q/c\right)\left(\vec{v}_{+}-\vec{v}_{-}\right)\times\vec{B}\left(\vec{r}_{0},t\right)$,
where $\vec{v}_{+}=d\vec{r}_{+}/dt$ is the velocity of the
positive charge $q$ of the dipole (and similarly for
$\vec{v}_{-}$). Thus, the magnetic force can be written as
$\frac{1}{c}\frac{d\vec{p}}{dt}\times\vec{B}$. Using the vector
identity
$\left(\vec{p}\cdot\nabla\right)\vec{E}=\text{grad}\left(\vec{p}\cdot\vec{E}\right)-\vec{p}\times\text{rot}\vec{E}$,
one can write the total force as \cite{gord}
\begin{equation}
\vec{f}=\text{grad}\left(\vec{p}\cdot\vec{E}\right)+\frac{1}{c}\frac{d}{dt}\left(\vec{p}\times\vec{B}\right).\label{f1}
\end{equation}

In our problem the dipole moment and the fields are fluctuating
quantities and, since the fluctuations are stationary, the last
term in Eq (\ref{f1}) disappears upon averaging. We are left with
the gradient term
\begin{equation}
\left\langle \vec{f}\right\rangle
\equiv\vec{F}=\nabla_{\vec{r}}\left\langle
\vec{p}\left(\vec{r}_{0},t\right)\cdot\vec{E}\left(\vec{r},t\right)\right\rangle
,\label{f2}
\end{equation}
where, again, setting $\vec{r}=\vec{r}_{0}$ after differentiation
is implied. This equation holds also for a dipole in motion and it
serves as the starting point for calculating the
fluctuation-induced forces \cite{intra,ded}.

We now consider an atom, or a nanoparticle, or any entity with
polarizability $\alpha\left(\omega\right)$ and size smaller than
the relevant wavelength of the electromagnetic field (we use below
the generic term ``particle''). We allow for the particle
temperature $T_{p}$ to be different from the sample temperature
$T_{L}$. The particle is placed at a distance $z_{0}$ above the
sample surface (see Fig. 2 for a schematic setup).

\begin{figure}[H]
\centering{}\includegraphics[scale=0.65]{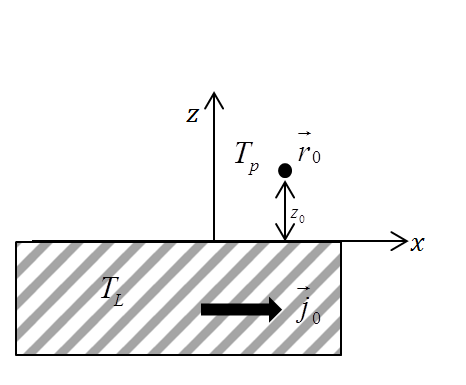}\caption{A small particle is placed at the point $\vec{r}_{0}=\left(x_{0},y_{0},z_{0}\right)$
above the sample surface. Both the sample and the particle are at
rest but a dc current with density $\vec{j}_{0}=en_{0}\vec{v}_{0}$
is flowing in the sample, in the $x$- direction. The temperature
of the sample, i.e., of its lattice, is $T_{L}$. The particle temperature
is $T_{p}$.}
\end{figure}

The force acting on the particle consists of two parts:
\begin{enumerate}
\item The fluctuating field emerging from the sample induces a
dipole moment in the particle. This emerging field, which is just
the field considered in the previous section, is often called
``free'' or ``spontaneous'' and will be designated as
$\vec{E}_{s}\left(\vec{r},t\right)$. Interaction of this field
with the dipole moment $\vec{p}_{i}\left(\vec{r}_{0},t\right)$
induced in the particle is responsible for the first part,
$\vec{F}_{1}$, of the force. \item The particle itself induces a
fluctuating electric field in the environment, due to the
spontaneous fluctuations of its dipole moment. We denote the
latter by $\vec{p}_{s}\left(\vec{r}_{0},t\right)$ and the
corresponding field by $\vec{E}_{i}\left(\vec{r},t\right)$ . This
field acts back on the particle, giving the second part,
$\vec{F}_{2}$, of the force. \end{enumerate}

Thus, Eq (\ref{f2}) splits into two
parts, containing respectively $\vec{p}_{i}\cdot\vec{E}_{s}$ and
$\vec{p}_{s}\cdot\vec{E}_{i}$.  Furthermore, since the particle
polarizability $\alpha\left(\omega\right)$ is frequency dependent,
one has to rewrite Eq (\ref{f2}) in frequency domain \cite{foot4}:
\begin{equation}
\vec{F}\left(\vec{r}_{0}\right)=\vec{F}_{1}+\vec{F}_{2}=\int_{-\infty}^{+\infty}\frac{d\omega}{2\pi}\alpha\left(\omega\right)\nabla_{\vec{r}}
\left\langle
\vec{E}_{s}\left(\vec{r}_{0}\right)\cdot\vec{E}_{s}^{*}\left(\vec{r}\right)\right\rangle
+\int_{-\infty}^{+\infty}\frac{d\omega}
{2\pi}\nabla_{\vec{r}}\left\langle
\vec{p}_{s}^{*}\left(\vec{r}_{0}\right)\cdot\vec{E}_{i}\left(\vec{r}\right)\right\rangle
_{\omega},\label{f3}
\end{equation}
where $\vec{p}_{i}\left(\vec{r}_{0},\omega\right)=\alpha\left(\omega\right)\vec{E}_{s}\left(\vec{r}_{0},\omega\right)$
has been used.


In the rest of this section we specialize to the case
$\nu\rightarrow 0$. Then the spectral density in the first term of
 Eq (\ref{f3}) is given, in somewhat different notations, in Eq (\ref{E'}), so
that $\vec{F}_{1}$ is obtained immediately, after applying
$\nabla_{\vec{r}}$ and setting $\vec{r}=\vec{r}_{0}$ at the end.
It is useful to split the integral over $\omega$ into two pieces:
from $-\infty$ to $0$ and from $0$ to $+\infty$ . Switching the
sign of the integration variables in the first piece, and using
the conditions
$\epsilon_{L}\left(-\omega\right)=\epsilon_{L}^{*}\left(\omega\right)$
, $\alpha\left(-\omega\right)=\alpha^{*}\left(\omega\right)$,
$\epsilon_{1}\left(-\omega,
-k_{x}\right)=\epsilon_{1}^{*}\left(\omega, k_{x}\right)$ , we
finally obtain the following expressions for the $x$ and $z$-
component of $\vec{F}_{1}$ (the $y$-component is zero) :
\begin{equation}
F_{1x}\left(z_{0}\right)=\frac{2\hbar}{\pi^{2}}\int_{0}^{\infty}d\omega\alpha^{\prime\prime}\left(\omega\right)
\epsilon_{L}^{\prime\prime}
\left(\omega\right)\coth\left(\frac{\hbar\omega}{2T_{L}}\right)\iintop_{-\infty}^{\infty}dk_{x}dk_{y}
\frac{qk_{x}} {\left|\epsilon_{1}\left(\omega,
k_{x}\right)+1\right|^{2}}e^{-2qz_{0}}\label{f4}
\end{equation}
\begin{equation}
F_{1z}\left(z_{0}\right)=-\frac{2\hbar}{\pi^{2}}\int_{0}^{\infty}d\omega\alpha^{\prime}\left(\omega\right)
\epsilon_{L}^{\prime\prime}\left(\omega\right)\coth\left(\frac{\hbar\omega}{2T_{L}}\right)\iintop_{-\infty}^
{\infty}dk_{x}dk_{y}
\frac{q^{2}}{\left|\epsilon_{1}\left(\omega,
k_{x}\right)+1\right|^{2}}e^{-2qz_{0}},\label{f5}
\end{equation}
where $\epsilon_{1}$ is defined in Eq (\ref{epsA}),
$\alpha^{\prime}\left(\omega\right)$,
$\alpha^{\prime\prime}\left(\omega\right)$ are the real and
imaginary parts of $\alpha\left(\omega\right)$, and
$q=\sqrt{k_{x}^{2}+k_{y}^{2}}$. Since both the particle and the
sample are at rest, it is natural that $\alpha$ and $\epsilon_{L}$
depend on $\omega$ but not on $k_x$ (no Doppler shift). The
Doppler shifted frequency $\omega_{-}=\omega-k_{x}v_{0}$ enters
only into the electronic part of $\epsilon_1$ which controls the
dynamics of the fluctuations.

We now turn to the second term in Eq (\ref{f3}). First, one needs
to compute the field $\vec{E}_{i}\left(\vec{r},\omega\right)$ induced
by the spontaneous fluctuations of the particle dipole moment $\vec{p}_{s}\left(\vec{r}_{0},\omega\right)$.
To this end we introduce the Green's function  $G\left(\vec{r},\vec{r}_{0},\omega\right)$ as
 a solution of the Poisson equation for a unit charge at point $\vec{r}_{0}$.
Then the electric potential created at point $\vec{r}$ by the ``point
dipole''  $\vec{p}_{s}$ and the electric field of that dipole are given
by
\begin{equation}
\Phi_{i}\left(\vec{r},\omega\right)=\vec{p}_{s}\left(\vec{r}_{0},\omega\right)\cdot\nabla_
{\vec{r}_{0}}G\left(\vec{r},\vec{r}_{0},\omega\right)\;\;,\;\;\vec{E}_{i}\left(\vec{r},\omega\right)=
-\nabla_{\vec{r}}\Phi_{i}\left(\vec{r},\omega\right),\label{f6}
\end{equation}
so that
\begin{equation}
\left\langle \vec{p}_{s}^*\left(\vec{r}_{0}\right)\cdot\vec{E}_{i}\left(\vec{r}\right)\right\rangle _{\omega}=-\left\langle p_{s\alpha}^*\left(\vec{r}_{0}\right)p_{s\beta}\left(\vec{r}_{0}\right)\right\rangle _{\omega}\frac{\partial^{2}}{\partial r_{\alpha}\partial r_{0\beta}}G\left(\vec{r},\vec{r}_{0},\omega\right).\label{f7}
\end{equation}
where $\alpha,\beta$ labels the components and summation over indices is
implied. The Green's function can be written as
\begin{equation}
G\left(\vec{r},\vec{r}_{0},\omega\right)=\int\frac{d^{2}q}{\left(2\pi\right)^{2}}e^{iq\left(\vec{\rho}-\vec{\rho}_{0}
\right)}g\left(z, z_{0}, \vec{q}, \omega\right)\label{f8}
\end{equation}
where $g\left(z, z_{0}, \vec{q}, \omega\right)$ satisfies
\begin{equation}
\epsilon_{1}\left(\omega,
k_{x}\right)\left(-\frac{\partial^{2}}{\partial z^{2}}+q^{2}\right)g\left(z, z_{0}, \vec{q}, \omega\right)=4\pi\delta\left(z-z_{0}\right)\label{f9}
\end{equation}
This is essentially the same as  Eq (\ref{green}), with $\epsilon\left(\omega, k_{x}\right)$
instead of $\epsilon\left(\omega\right)$, but now we have both the
source and the observation point above the sample surface, i.e., $z,z_{0}>0$.

There is an apparent difficulty here, namely: $G\left(\vec{r},\vec{r}_{0},\omega\right)$, being a response to a point source, is singular at $\vec{r}=\vec{r}_{0}$,
and so is $g\left(z,z_{0},\vec{q},\omega\right)$ for $z=z_{0}$.
Since in Eq (\ref{f3}) we have to further differentiate the expression
in Eq (\ref{f7}) with respect to $\vec{r}$ and then set $\vec{r}=\vec{r}_{0}$,
we end up with a meaningless singular expression. This problem is
well known in the theory of Casimir - Lifshitz forces (see, e.g., \cite{bim}) and the remedy
is to make the standard subtraction of the vacuum Green's function $G_{0}\left(\vec{r},\vec{r}_{0},\omega\right)$,
Thus, the physical Green's function is $\tilde{G}\left(\vec{r},\vec{r}_{0},\omega\right)=G\left(\vec{r},\vec{r}_{0},\omega\right)-G_{0}\left(\vec{r},\vec{r}_{0},\omega\right)$
or the Fourier transformed $\tilde{g}\left(z,z_{0},\vec{q},\omega\right)=g\left(z,z_{0},\vec{q},\omega\right)-g_{0}\left(z,z_{0},\vec{q},\omega\right)$,
where $g_{0}$ is obtained from $g$ by replacing $\epsilon\left(\omega, k_{x}\right)$
by unity. The result is:
\begin{equation}
\tilde{g}\left(z,z_{0},\vec{q},\omega\right)=-\frac{2\pi}{q}e^{-q\left(z+z_{0}\right)}\Gamma_{1}\left(\omega, k_{x}\right), \;\;\;\;\;\;\;\;
\Gamma_{1}\left(\omega, k_{x}\right)=\frac{\epsilon_{1}\left(\omega, k_{x}\right)-1}{\epsilon_{1}\left(\omega, k_{x}\right)+1}\label{f10}
\end{equation}
The last piece of information that we need to complete the calculation
is the expression for the spectral density \cite{gin}
\begin{equation}
\left\langle p_{s\alpha}^*\left(\vec{r}_{0}\right)p_{s\beta}\left(\vec{r}_{0}\right)\right\rangle _{\omega}=\delta_{\alpha\beta}\hbar\alpha^{\prime\prime}\left(\omega\right)\coth\left(\frac{\hbar\omega}{2T_{p}}
\right)\label{f11}
\end{equation}
where an isotropic particle, with temperature $T_{p},$ has been assumed.
Putting all pieces together, and using $\Gamma_{1}\left(-\omega, -k_{x}\right)= \Gamma_{1}^{*}\left(\omega, k_{x}\right)$ we obtain the following expression
for the components of $\vec{F}_{2}$:
\begin{equation}
F_{2x}\left(z_{0}\right)=-\frac{\hbar}{\pi^{2}}\int_{0}^{\infty}d\omega\alpha^{\prime\prime}\left(\omega\right)
\coth\left(\frac{\hbar\omega}{2T_{p}}\right)\iintop_{-\infty}^{\infty}dk_{x}dk_{y}
\Gamma_{1}^{\prime\prime}\left(\omega, k_{x}\right)qk_{x}e^{-2qz_{0}}\label{f12},
\end{equation}

\begin{equation}
F_{2z}\left(z_{0}\right)=-\frac{\hbar}{\pi^{2}}\int_{0}^{\infty}d\omega\alpha^{\prime\prime}\left(\omega\right)
\coth\left(\frac{\hbar\omega}{2T_{p}}\right)\iintop_{-\infty}^{\infty}dk_{x}dk_{y}
\Gamma_{1}^{\prime}\left(\omega, k_{x}\right)q^{2}e^{-2qz_{0}}\label{f13},
\end{equation}
where $\Gamma_{1}^{\prime}\left(\omega, k_{x}\right)$ and $\Gamma_{1}^{\prime\prime}\left(\omega, k_{x}\right)$ are the real and imaginary
 parts of $\Gamma_{1}\left(\omega, k_{x}\right)$, Eq
 (\ref{f10}). The total force acting on the particle is the sum of
 $\vec{F}_{1}$ and $\vec{F}_{2}$. In the next section we consider
 some specific examples.

\section{Fluctuation - Induced Forces: Summary, Discussion and Examples}

Let us summarize our general results for the fluctuation-induced
forces, acting on a particle in the presence of drifting mobile
carriers in the medium. The medium is described by the Drude model
with drift, Eq (\ref{epsilons}), and two opposite limits were
considered:

\uline{Model 1}: The dominant contribution to the spontaneous random sources comes from the lattice and dissipation in the electron plasma is neglected.
The dielectric function of the model is $\epsilon_{1}\left(\omega, k_{x}\right)$,
Eq (\ref{epsA}).

\uline{Model 2}: The dominant contribution comes from the electron plasma and dissipation in the lattice is neglected. The
dielectric function of the model is $\epsilon_{2}\left(\omega, k_{x}\right)$, Eq
(\ref{epsB}).

The main reason for introducing these two limits, rather than
dealing with the general model, Eq (\ref{epsilons}), is that the
frequency of the fluctuation sources originating in the drifting
plasma is Doppler shifted with respect to those originating in the
lattice. Thus, while it is easy to write down the spectral density
of the sources for the general case (this is just the sum of Eqs (\ref{corrA}) and (\ref{corrB})), it would make the resulting
expressions for the forces even more cumbersome and more difficult
to analyze. We therefore prefer to clarify the basic physics of
the problem using the two limiting models.

Let us first return to "Model 1". Since
$\epsilon_{L}^{\prime\prime}(\omega) =
\epsilon_{1}^{\prime\prime}\left(\omega, k_{x}\right)$, we have
\begin{equation}
\frac{\epsilon_{L}^{\prime\prime}(\omega)}{\left|\epsilon_{1}\left(\omega,
k_{x}\right)+1\right|^{2}} =
\frac{1}{2}Im[\frac{\epsilon_{1}\left(\omega,
k_{x}\right)-1}{\epsilon_{1}\left(\omega,
k_{x}\right)+1}]=\frac{1}{2}
\Gamma_{1}^{\prime\prime}\left(\omega, k_{x}\right).\label{F1}
\end{equation}
This identity enables one to write Eqs (\ref{f4}, \ref{f5}) in
terms of $\Gamma_{1}^{\prime\prime}\left(\omega, k_{x}\right)$.
Adding to Eqs (\ref{f4}, \ref{f5}) their counterparts in Eqs
(\ref{f12}, \ref{f13}) gives the final expressions for the
components of the total force in "Model 1":
\begin{equation}
\begin{array}{c}
F_{x}\left(z_{0}\right)=\frac{\hbar}{\pi^{2}}\int_{0}^{\infty}d\omega\alpha^{\prime\prime}\left(\omega\right)
\left[\coth\left(\frac{\hbar\omega}{2T_{L}}\right) -
\coth\left(\frac{\hbar\omega}{2T_{p}}\right)\right]\\
\iintop_{-\infty}^{\infty}dk_{x}dk_{y}
\Gamma_{1}^{\prime\prime}\left(\omega,
k_{x}\right)qk_{x}e^{-2qz_{0}}.\label{F2}
\end{array}
\end{equation}

\begin{equation}
\begin{array}{c}
F_{z}\left(z_{0}\right)=-\frac{\hbar}{\pi^{2}}\int_{0}^{\infty}d\omega\iintop_{-\infty}^{\infty}dk_{x}dk_{y}
[\alpha^{\prime}\left(\omega\right)\coth\left(\frac{\hbar\omega}{2T_{L}}\right)\Gamma_{1}^{\prime\prime}\left(\omega,
k_{x}\right)\\
+ \alpha^{\prime\prime}\left(\omega\right)
\coth\left(\frac{\hbar\omega}{2T_{p}}\right)
\Gamma_{1}^{\prime}\left(\omega,
k_{x}\right)]q^{2}e^{-2qz_{0}}.\label{F3}
\end{array}
\end{equation}
Let us clarify a bit these expressions, starting with Eq
(\ref{F3}). This normal-to-surface component is the generalization
of the standard, equilibrium Lifshitz force between a particle and
medium \cite{gin}. The generalization includes the effect of
carrier drift in the medium and it allows for different
temperatures of the medium and the particle. The first part of the
force, proportional to $\coth\left(\hbar\omega/2T_{L}\right)$, is
due to the fluctuating field in the medium acting on the particle.
The particle itself  is "passive", hence
$\alpha^{\prime}\left(\omega\right)$. In the second part,
proportional to $\coth\left(\hbar\omega/2T_{p}\right)$, the
fluctuating field originates in the particle and, after being
"reflected" from the medium, acts back on the particle. Here the
medium is passive, hence $\Gamma_{1}^{\prime}$. Note that both
$\coth$-functions have in their argument the unshifted frequency
$\omega$. This is because in "Model 1" the spontaneous fluctuating
sources of the medium reside in the lattice, which is at rest in
the laboratory system (the particle is at rest as well). The
Doppler shifted frequency $\omega_{-}$ appears only in
$\Gamma_{1}$ which contains information on the effect of the drift
on fluctuation  dynamics.

The structure of Eq (\ref{F2}) is different. This equation
describes the dissipative "drag" force, due to the current flow in
the medium.
 For this force to
exist both $\alpha^{\prime\prime}\left(\omega\right)$ and
$\Gamma_{1}^{\prime\prime}$ (i.e.,
 $\epsilon_{L}^{\prime\prime}$)
must differ from zero. However, the "active" part of the system
can be distinguished from the "passive" one by looking at the
argument of the $\coth$. The first term in Eq (\ref{F2}),
proportional to $\coth\left(\hbar\omega/2T_{L}\right)$, is due to
the random sources in the medium, i.e., the medium is the emitter
while the particle is the absorber (and vice versa for the second
term).

To switch to "Model 2" the following replacements are required in Eqs (\ref{F2},\ref{F3}):
The lattice temperature $T_L$ is replaced by the temperature of the electron plasma $T_{el}$  and $\Gamma_1$ is
 changed to $\Gamma_2$, which is defined in Eq (\ref{f10}), with subscript $2$ instead of {1}.
  Furthermore, since the spontaneous sources in the medium now originate in the drifting plasma,
   the frequency in the argument of the corresponding $\cosh$-function should be Doppler shifted.
    Thus, the counterparts of the Eqs (\ref{F2},\ref{F3}) for "Model 2" read as:
\begin{equation}
\begin{array}{c}
F_{x}\left(z_{0}\right)=\frac{\hbar}{\pi^{2}}\int_{0}^{\infty}d\omega\alpha^{\prime\prime}\left(\omega\right)
\iintop_{-\infty}^{\infty}dk_{x}dk_{y}
\left[\coth\left(\frac{\hbar\omega_{-}}{2T_{el}}\right) -
\coth\left(\frac{\hbar\omega}{2T_{p}}\right)\right]\\
\Gamma_{2}^{\prime\prime}\left(\omega,
k_{x}\right)qk_{x}e^{-2qz_{0}}.\label{F4}
\end{array}
\end{equation}

\begin{equation}
\begin{array}{c}
F_{z}\left(z_{0}\right)=-\frac{\hbar}{\pi^{2}}\int_{0}^{\infty}d\omega\iintop_{-\infty}^{\infty}dk_{x}dk_{y}
[\alpha^{\prime}\left(\omega\right)\coth\left(\frac{\hbar\omega_{-}}{2T_{el}}\right)\Gamma_{2}^{\prime\prime}
\left(\omega,
k_{x}\right)\\
+ \alpha^{\prime\prime}\left(\omega\right)
\coth\left(\frac{\hbar\omega}{2T_{p}}\right)
\Gamma_{2}^{\prime}\left(\omega,
k_{x}\right)]q^{2}e^{-2qz_{0}}.\label{F5}
\end{array}
\end{equation}

The above expressions for the forces resemble those obtained in
the literature for
 the problem of
non-contact friction, experienced by a particle moving above a medium at rest
\cite{volo, kya, ded} (or, alternatively, the problem of "drag"
exerted on the particle by a moving medium). Our problem, however,
is different and so are the results. Since in our setup   the
plasma component is moving with respect to the lattice, the
dielectric function governing dynamics of the fluctuations is in general a
complicated function of $\omega$ and $k_{x}$. In addition, the
frequency dependence of the random sources in the drifting plasma
is different (Doppler shifted) with respect to those in the
stationary lattice. Due to these factors the results are sensitive
to the details of the model and can be quite diverse. For instance,
in "Model 1" there are no "drag" at all, if the temperature of the
sample and the particle are equal, see Eq (\ref{F2}) with
$T_{L}=T_{p}$. This is because, as has been mentioned above, in "Model 1" the random sources, both in the medium and in the particle, are at rest. Only the dielectric function $\epsilon_{1}\left(\omega,
k_{x}\right)$ is affected by the drift. Therefore the situation is the same as in equilibrium, but with a modified, $k_x$-dependent dielectric function  of the medium.

To obtain specific results we need an explicit expression for  the
particle susceptibility $\alpha\left(\omega\right)$. We shall use
the most simple, "generic" expression applicable to a two-level
system:
\begin{equation}
\alpha\left(\omega\right) =
\frac{\alpha\left(0\right)\omega_{0}^{2}}{\omega_{0}^{2}-\omega^{2}-i\omega\eta},\label{F6}
\end{equation}
where $\omega_{0}$ is the resonance frequency of the excitation
and $\eta$ is the decay rate. This expression is valid for an atom
or a molecule when a single excitation is of importance. It is
also applicable to a metallic or semiconducting (spherical)
particle, in which case $\alpha\left(0\right)$ is equal to the
cube of the radius of the sphere and
$\omega_{0}=\tilde\omega_{p}/\sqrt{3}$ is the frequency of the
localized surface plasmon \cite{mai}. (Here $\tilde\omega_{p}$ is
the plasma frequency of the material of the particle). For a
dielectric nanoparticle one may have some phonon mode or a
phonon-polariton, instead of a plasmon. The value of $\omega_{0}$
depends on the nature of the particle and can vary over a few
orders of magnitude, say, between $10^{12}$ and $10^{16}
sec^{-1}$.

In the weak dissipation (small $\eta$) limit the imaginary part of
$\alpha\left(\omega\right)$ is often approximated as
\begin{equation}
\alpha^{\prime\prime}\left(\omega\right) =
\frac{\alpha\left(0\right)\omega_{0}^{2}\omega\eta}{(\omega_{0}^{2}-\omega^{2})^2
+(\omega\eta)^2}\Rightarrow
 \frac{\pi}{2}\alpha\left(0\right)\omega_{0}\delta(\omega-\omega_{0}), \;\;\;\;\;\;\;\;(\omega >0).
\label{F7}
\end{equation}
However, one should keep in mind that, when
$\alpha^{\prime\prime}\left(\omega\right)$ is integrated with some
function  of frequency $f(\omega)$, the "$\delta$-function
approximation" is valid only if $f(\omega_0)$ is not negligibly
small. Otherwise the integral will be dominated not by the peak of
the Lorenzian in Eq (\ref{F7}) but by some other region of
frequencies where  $f(\omega)$ is significant (albeit the
Lorenzian is small). Below we shall encounter a situation where
the integral is dominated by small frequencies and,
correspondingly, the low-frequency expansion
\begin{equation}
\alpha^{\prime\prime}\left(\omega\right) =
\frac{\alpha\left(0\right)\omega\eta}{\omega_{0}^{2}} \label{F8}
\end{equation}
will be used.

The same remark applies to $\Gamma_{1}^{\prime\prime}\left(\omega,
k_{x}\right)$, defined in Eq (\ref{F1}) (and similarly for
$\Gamma_{2}^{\prime\prime}\left(\omega, k_{x}\right)$,  with
$\epsilon_{2}\left(\omega, k_{x}\right)$ instead of
$\epsilon_{1}\left(\omega, k_{x}\right)$). In the small
dissipation limit, the general expression can be approximated, in
some cases, by the $\delta$-function
\begin{equation}
\Gamma^{\prime\prime}\left(\omega_{-}\right)=\pi(1-C)
\delta(\omega_{-}^2-\omega_{sp}^2), \;\;\;\;\;\;\;\;
\omega_{sp}=\frac{\omega_{p}}{\sqrt{\epsilon_{L}^{\prime}+1}},
\;\;\;\;\;\;\;\;C\equiv
\frac{(\epsilon_{L}^{\prime}-1)}{(\epsilon_{L}^{\prime}+1)}.\label{F9}
\end{equation}
It is assumed here that the relevant frequencies are far from the
resonant frequencies of the  lattice and $C$ can be treated as a
constant, hence $\omega$ and $k_x$ are combined into a single
argument $\omega_{-}$. In the $\delta$-function approximation
there is no difference between $\Gamma_{1}$ and $\Gamma_{2}$, so
that the subscript has been removed. Note that Eq (\ref{F9}) does
not explicitly contain $\epsilon_{L}^{\prime\prime}$ or $\nu$
although some dissipation, albeit infinitely small, is essential.
Since, however, in reality the dissipation is finite, the
$\delta$-function approximation has its limitations and, in
particular, below we shall need the small frequency approximation
\begin{equation}
\Gamma_{2}^{\prime\prime}\left(\omega_{-}\right)=
\frac{2\nu\omega_{-}}{\omega_{p}^2}. \label{F10}
\end{equation}

 We are now in a position to work
out some examples of the drift effect on fluctuation induced
forces. The most interesting effect is the appearance of the
aforementioned drag force.

\subsection{Drag force in "Model 1"}

In this model the drag force on a particle appears only if $T_{p}$
and $T_{L}$ are different, see Eq (\ref{F2}). Note that if $T_{p}$ and $T_{L}$ are reversed, the force changes
sign, i.e., drag (force in the direction of the current) turns
into "anti-drag" (force in the opposite direction)\cite{foot5}.
 Let us calculate the force $F_x$ using  the $\delta$-approximation for $\alpha^{\prime\prime}\left(\omega\right)$, Eq (\ref{F7}).
 This is justified because the integral is dominated by the peak of the Lorenzian. The $\delta$-function
takes care of the integral over $\omega$ in Eq (\ref{F2}), and we
have to address the integral over $k_{x}, k_{y}$, with
$\Gamma_{1}^{\prime\prime}\left(\omega_{0}, k_{x}\right)$. The
latter quantity is defined in (\ref{F1}). Since
$\epsilon_{L}^{\prime\prime}$ is a small number,
$\Gamma_{1}^{\prime\prime}\left(\omega_{0}, k_{x}\right)$ has a
sharp maximum when $\epsilon_{1}^{\prime}\left(\omega_{0}, k_{x}\right)+1 =
0$. This happens at $k_x=k_x^{\pm}=(\omega_{0} \pm
\omega_{sp})/v_0$. One can try to approximate the Lorenzian function in
Eq (\ref{F1}) by the
$\delta$-function, Eq (\ref{F9}), thus obtaining
\begin{equation}
\Gamma_{1}^{\prime\prime}\left(\omega_{0},
k_{x}\right)=\frac{\pi}{\epsilon_{L}^{\prime}+1}\frac{\omega_{sp}}{v_0}[\delta(k_x-k_x^+)
 +\delta(k_x-k_x^-)].
 \label{D12}
\end{equation}
In order to see how good is this approximation one must keep in
mind that, due to the exponential factor, the integrand in Eq
(\ref{F2}) has a sharp cutoff at $k_x\sim 1/z_0$, hence the
$\delta$-approximation will be justified only if at least one of
the roots $k_x^{\pm}$ is below
the cutoff- otherwise the contribution from the peak of
$\Gamma_{1}^{\prime\prime}\left(\omega_{0}, k_{x}\right)$ is
exponentially small (we assume here that both roots are positive).
The $\delta$-approximation is always justified for sufficiently
large $v_0$ but the precise criterion depends on the values of
$\omega_{sp}$, $\omega_{0}$ and $z_0$. For an atom $\omega_{0}$ is
typically much  larger than $\omega_{sp}$ of the semiconducting
medium but for a large molecule or a nanoparticle  (dielectric or
semiconducting) the two frequencies can be of the
same order. We assume that $\omega_{0}$ is few times larger than
$\omega_{sp}$ and obtain the condition $\omega_{0}z_0/v_0 \ll 1$
for the validity of the $\delta$-approximation. The force $F_x$
is then estimated from (\ref{F2}) as
\begin{equation}
F_x\sim \frac{\hbar
\alpha(0)}{\epsilon_{L}^{\prime}+1}\omega_{sp}\left(\frac{\omega_{0}}{v_0z_0}\right)^2
\left[\coth\left(\frac{\hbar\omega_{0}}{2T_{L}}\right) -
\coth\left(\frac{\hbar\omega_{0}}{2T_{p}}\right)\right],\;\;\;\;\;\;\;\;
(v_0\gg \omega_{0}z_0).
 \label{D13}
\end{equation}
In this regime $F_x$ drops as $v_0^{-2}$ under increase of the
drift velocity. It achieves its maximum value for $v_0\sim
\omega_{0}z_0$, at which point the $\delta$-approximation breaks
down. For a hot medium, ($\hbar\omega_{0}/2T_{L})<<1$, and a "cold
particle", ($\hbar\omega_{0}/2T_{p})>>1$, this maximum value is of
the order of
$\alpha_0(0)\omega_{sp}T_L/\omega_{0}z_0^4(\epsilon_{L}^{\prime}+1)$
which is comparable with the equilibrium Casimir-Lifshitz force.

In the opposite case of small drift velocities the
 $\delta$-approximation breaks down and the integral is dominated
by small $k_x$.
$\Gamma_{1}^{\prime\prime}\left(\omega_0-k_xv_0\right)$ should
then be expanded near the point $\omega_0$ with respect to
$k_xv_0$. The expansion has a linear term, unless
$\omega_{sp}$=$\omega_{0}$ when the first correction is quadratic in $v_0$. We
assume to be well away from this point, taking $\omega_{0}$ few
times larger than $\omega_{sp}$.
The first contribution to $F_x$ comes then from the linear term which,
after substitution into (\ref{F2}) and integration over $k_x,
k_y$, yields
\begin{equation}
F_x=\frac{3\hbar v_0
\alpha(0)}{z_0^5}\frac{\epsilon_{L}^{\prime\prime}}{\left(\epsilon_{L}^{\prime}+1\right)^2}\left(\frac{\omega_{sp}}{\omega_{0}}\right)^2
\left[\coth\left(\frac{\hbar\omega_{0}}{2T_{L}}\right) -
\coth\left(\frac{\hbar\omega_{0}}{2T_{p}}\right)\right],\;\;\;\;\;\;\;\;
(v_0\ll \omega_{0}z_0). \label{D14}
\end{equation}
Note that the two expressions, Eqs (\ref{D13}) and (\ref{D14}), do
not match at $v_0\sim \omega_{0}z_0$ which means that there is an
intermediate region where $F_x$ sharply increases, interpolating
between the small and large velocity limits. The overall behavior
of $F_x$, as a function of $v_0$, is
schematically scetched in Fig. 3. If one takes $\omega_{0} \sim
10^{12} sec^{-1}$ and $z_0 \sim 10nm$, then the critical drift velocity
 for which the maximal value of the force is reached, is $v_{0c} \sim
10^6 cm/sec$.
Although this is comparable to a typical saturation velocity in semiconductors, reaching the maximum force value and, morover, observing the $1/v_0^2$ decay might well be unrealistic. In addition to the very small  $\omega_{0}$ and very large $v_0$ needed for such observation, it is not at all clear that under such extreme conditions the electron plasma can be characterized by a temperature.
\begin{figure}[H]
\centering{}\includegraphics[scale=0.4]{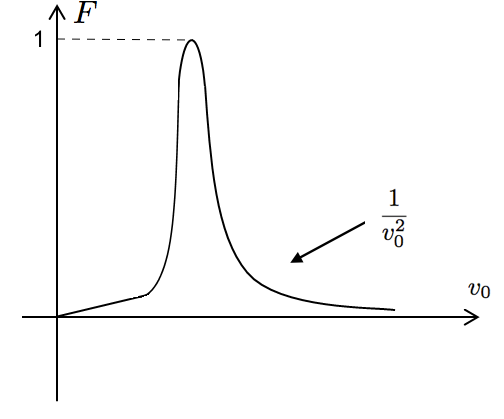}\caption{A qualitative plot of the drag force in "Model 1", as a function of $v_0$. The force is given in units of its maximal value, see text.  For small  $v_0$ the behavior is linear. The large-$v_0$ asymptotic is  $1/v_0^2$. }
\end{figure}

\subsection{Drag force in "Model 2"}
The expression for the force is given in Eq (\ref{F4}). Due to the
Doppler shifted frequency in the argument of the first $\coth$,
the force exists also when the medium and the particle have equal
temperatures, $T_L=T_p\equiv T$, and we concentrate on this case.
The most interesting limit is $T\rightarrow 0$ ("quantum drag").
In this limit the difference between the two $\coth$ functions in
Eq (\ref{F4}) is equal to $-2$ for $0<\omega<k_xv_0$ and it is
zero otherwise (recall that $\omega$ must be positive). Thus,
Eq (\ref{F4}) reduces to

\begin{equation}
F_{x}\left(z_{0}\right)=-2\frac{\hbar}{\pi^{2}}
\iintop_{-\infty}^{\infty}dk_{x}dk_{y}\int_{0}^{k_xv_0}d\omega\alpha^{\prime\prime}\left(\omega\right)\Gamma_{2}^{\prime\prime}\left(\omega_{-}
\right) qk_{x}e^{-2qz_{0}}.\label{D15}
\end{equation}

 One might be tempted to approximate the Lorenzian
$\alpha^{\prime\prime}\left(\omega\right)$ by the
$\delta$-function in Eq (\ref{F7}). This. however. is possible
only if the upper limit in the integral over $\omega$ is larger
than the center $\omega_{0}$ of the Lorenzian peak. Since the
relevant values of $k_x$ are smaller than $1/z_0$, we arrive again
to the parameter ($v_0/\omega_{0}z_0)\equiv \gamma$. Only if this
parameter is large can one justify the "$\delta$-approximation".

For small value of $\gamma$ the upper limit in the integral over
frequency is smaller than $\omega_{0}$ and one should use the
low-frequency expressions for
$\alpha^{\prime\prime}\left(\omega\right)$ and
$\Gamma_{2}^{\prime\prime}\left(\omega_{-}\right)$,  Eqs
(\ref{F8})) and (\ref{F10})) respectively.
Note that in the integration region over the frequency,
$\omega_{-}$ is negative, and so is
$\Gamma_{2}^{\prime\prime}\left(\omega_{-}\right)$. The integral
over $\omega$ is proportional to $v_0^3$ and the force is of the
order of
\begin{equation}
F_x\sim\frac{\hbar \alpha(0)\eta\nu
v_0^3}{z_0^7\omega_0^2\omega_p^2}\;\;\;\;\;\;\;\; (v_0\ll
\omega_{0}z_0). \label{D16}
\end{equation}

For large $\gamma$ the main contribution to the integral over
frequencies in Eq (\ref{D15}) comes from high frequencies,
$\omega\simeq \omega_0$, and the $\delta$-approximation for
$\alpha^{\prime\prime}\left(\omega\right)$ is valid. Thus, first
one integrates over $\omega$ and then over $k_x$, using the
$\delta$-approximation for
$\Gamma_{2}^{\prime\prime}\left(\omega_0-k_xv_0\right)$. Due to
the restriction $\omega_0<k_xv_0$,
$\Gamma_{2}^{\prime\prime}\left(\omega_0-k_xv_0\right)$ is now
negative and its argument has only one root, namely,
$k^{+}=(\omega_{0}+\omega_{sp})/v_0$. Assuming again that
$\omega_0$ is few times larger than $\omega_{sp}$, we arrive to a
simple estimate
\begin{equation}
F_x(z_0)\sim \frac{\hbar
\alpha(0)}{\epsilon_{L}^{\prime}+1}\omega_{sp}\left(\frac{\omega_{0}}{v_0z_0}\right)^2
,\;\;\;\;\;\;\;\; (v_0\gg \omega_{0}z_0).
 \label{D17}
\end{equation}
The fundamental difference between this expression and its
counterpart in "Model 1", Eq (\ref{D13}), is that  Eq (\ref{D17})
was derived in the zero-temperature limit, when  Eq (\ref{D13})
(as generally for equal temperatures of the particle and the
medium) is zero. The qualitative behavior of $F_x$ in  Eq
(\ref{D17}), as a function of $v_0$, is similar to that shown in
Fig.3, although the initial slope is less steep (proportional to
$v_0^3$ instead of being linear). The maximal value of the force,
$F_x(z_0)\sim
 \hbar \alpha(0)\omega_{sp}/z_0^4$, is achieved for $v_0\sim
\omega_{0}z_0$ (in this estimate we take $\omega_0$ to be few
times larger than $\omega_{sp}$ and assume
$\epsilon_{L}^{\prime}\sim 1$). This force is of the same
magnitude as the usual Casimir-Lifshitz attraction force between a
particle and a medium, in equilibrium.

\subsection{Effect of drift on $F_{z}$}

Unlike the lateral force $F_x$, the normal (Casimir-Lifshitz)
force exists already in the equilibrium. This force, however, is
affected by the drift of the mobile carriers. To concentrate
exclusively on the effect of drift we take $T_L=T_p\equiv T$ and
consider "Model 1", Eq (\ref{F3}). In this case the integral over
frequencies can be reduced to a Matsubara sum, in spite of the
fact that the system is not in equilibrium. Indeed,
$\coth\left(\frac{\hbar\omega}{2T}\right)$ becomes a common factor
for both terms in Eq (\ref{F3}) and they can be combined into an
expression containing
$Im\left[\alpha\left(\omega\right)\Gamma_{1}\left(\omega,
k_{x}\right)\right]$ which results in a Matsubara sum
\begin{equation}
F_{z}\left(z_{0}\right)=-\frac{\hbar}{\pi^{2}}\frac{2\pi
 T}{\hbar}Re{\sum_{n=0}^{\infty}}'
\iintop_{-\infty}^{\infty}dk_{x}dk_{y}\alpha\left(i\zeta_n\right)\Gamma_{1}\left(i\zeta_n,
k_{x}\right)q^2e^{-2qz_{0}},\label{Z1}
\end{equation}
where $\zeta_n = 2\pi Tn/\hbar$ ($n=0,1,...$) and the prime on
$\sum$ indicates that the $n=0$ term should be taken with a factor
$1/2$. In equilibrium $\alpha\left(i\zeta_n\right)$ and
$\Gamma_{1}\left(i\zeta_n\right)$ are real so that the sign "Re"
in front of the sum becomes redundant. In the presence of drift,
however, $\Gamma_1$ acquires $k_x$-dependence and becomes complex
on the imaginary frequency axis. Neglecting small dissipation, we have
\begin{equation}
\alpha\left(i\zeta\right) =
\frac{\alpha\left(0\right)\omega_{0}^{2}}{\omega_{0}^{2}+\zeta^{2}},
\;\;\;\;\;\;\;\;\; \Gamma(i\zeta, k_x)=
\frac{C(i\zeta-k_xv_0)^2-\omega_{sp}^2}{(i\zeta-k_xv_0)^2-\omega_{sp}^2},\label{Z2}
\end{equation}
where again $C$, which is generally a function of frequency, is treated here as a constant.

In the low-T limit, i.e., $T<<\omega_{0}, \omega_{sp}$ (these are
frequency scales on which $\alpha$ and $\Gamma$ changes
significantly). the sum can be replaced by an integral according
to the rule $\sum_{n}^{\prime}
\alpha\left(i\zeta_n\right)\Gamma_{1}\left(i\zeta_n,
k_{x}\right)=\frac{\hbar}{2\pi
 T}\int_{0}^{\infty}d\zeta\alpha\left(i\zeta\right)
\Gamma_{1}(i\zeta,
 k_{x})$, i.e.,
\begin{equation}
F_{z}\left(z_{0}\right)=-\frac{\hbar}{\pi^{2}}
Re\int_{0}^{\infty}d\zeta\alpha\left(i\zeta\right)\iintop_{-\infty}^{\infty}dk_{x}dk_{y}\Gamma_{1}\left(i\zeta,
k_{x}\right)q^2e^{-2qz_{0}}.\label{Z3}
\end{equation}
One can compute the correction to the force, due to carrier drift, by expanding $\Gamma_{1}\left(i\zeta_n,
k_{x}\right)$ in powers of $v_0$. The zero-order term corresponds to equilibrium, when $\Gamma_{1}\left(i\zeta_n,
k_{x}\right)$ does not depend on $k_x$ and
\begin{equation}
F_{z}^{0}\left(z_{0}\right)=-\frac{3\hbar}{4\pi
 z_{0}^4}\int_{0}^{\infty}d\zeta\alpha\left(i\zeta\right)\Gamma_{1}\left(i\zeta, 0\right)=-\frac{3\hbar}{8
z_{0}^4}\alpha\left(0\right)\omega_{0}\frac{C\omega_{0}+\omega_{sp}}{\omega_{sp}+\omega_{0}}
.\label{Z4}
\end{equation}

%
%
%
For $C=0$ this coincides with the well known expression for the attraction force between a "two-level atom" and a collisionless plasma \cite{gin}. The coefficient $C$ in  Eq (\ref{Z4}) accounts for the effect of the lattice.
The first order correction, i.e., the one linear in  $k_xv_0$,  does not contribute to the force. The second order corrrection \begin{equation}
\Delta\Gamma(i\zeta,k_x)=
(1-C)\frac{\omega_{sp}^2}{(\omega_{sp}^2+\zeta^{2})^2}(k_xv_0)^2
\frac{\omega_{sp}^2-3\zeta^{2}}{\omega_{sp}^2+\zeta^{2}}.
 \label{Z5}
\end{equation}
contributes to Eq (\ref{Z3}) the term
\begin{equation}
\Delta F_{z}\left(z_{0}\right)=-\frac{15}{16
z_0^6}(1-C)\hbar\alpha\left(0\right)
\frac{\omega_{0}\omega_{sp}v_0^2}{(\omega_{0} + \omega_{sp})^3}.
 \label{Z6}
\end{equation}
Assuming that $C$ is not close to $1$ and taking, as before,
$\omega_{0}$ to be few times larger than $\omega_{sp}$, one
recovers the same condition $v_0\ll \omega_{0}z_0$ for the
validity of the expansion.

Eq (\ref{Z6}) can be derived directly from Eq (\ref{F3}), without
using the Matsubara representation, although the latter is more
flexible when it comes to non-negligible dissipation and arbitrary
temperatures. The transformation of the expression in Eq
(\ref{F3}) to the Matsubara sum was possible because in "Model 1"
(and for $T_L = T_p$) the spontaneous fluctuation sources are in
equilibrium at the same temperature, only the (noiseless) plasma
component is in motion. This is not the case for "Model 2", where
the sources originating in the moving plasma are Doppler shifted,
so one has to work directly with the expression in Eq (\ref{F5}).
For small drift velocities and in the weak dissipation limit, when
the $\delta$-approximation for $\alpha^{\prime\prime}$ and
$\Gamma_{2}^{\prime\prime}$ can be used, the calculation is quite
straightforward and will not be pursued here. Instead, we briefly
discuss the case when the sample and the particle have different
temperatures but there is no drift. Then, for negligible
dissipation, Eqs (\ref{F3}) and  (\ref{F5}) become identical and
the result is the same for either model:
\begin{equation}
 F_{z}\left(z_{0}\right)=-\frac{3\hbar}{8z_0^4}\frac{\alpha\left(0\right)\omega_{0}}{\omega_{0}^{2}-\omega_{sp}^{2}}
\left[\omega_{0}\omega_{sp}(1-C)
\coth\left(\frac{\hbar\omega_{sp}}{2T_{L}}\right) +
\left(C\omega_{0}^2-\omega_{sp}^2\right)\coth\left(\frac{\hbar\omega_{0}}{2T_{p}}\right)\right].\label{Z7}
\end{equation}
This is a slight generalization of the result obtained in
\cite{bart} where $C=0$, i.e., $\epsilon_{L}^{\prime}=1$. This
latter case is appropriate for the free electron gas model, while
the expression (\ref{Z7}) includes the effect of the underlying
lattice. The constant $C$ can vary between $0$ and $1$, and for a
typical semiconductor, in  broad intervals of frequencies,
it can be few tenths or even close to $1$, so its effect is quite significant. The  ($C\omega_{0}^2$)-term in
(\ref{Z7}) can become the dominant one. For instance,
taking the low temperature limit, i.e., replacing the $\coth$-
factors by $1$, and assuming $\omega_{sp}<<\omega_{0}$, one
obtains $F_{z}=-(3\hbar/8z_0^4)\alpha\left(0\right)C\omega_{0}$.
This should be compared with
$F_{z}=-(3\hbar/8z_0^4)\alpha\left(0\right)\omega_{sp}$ for the
 electron gas model under the same conditions. The interesting
feature, pointed out already in \cite{bart}, is that, depending on
the parameters of the model, the force can be either repulsive or
attractive.

\section{Conclusion}
We have studied the fluctuation-induced forces acting on a small polarizable neutral particle (atom, molecule or a nanoparticle), located close to the surface of a conducting medium. It is shown that presence of a dc current (i.e., the mobile carrier drift) in the medium can have a significant effect on the forces. In particular, there appears a lateral force which can be  in the direction of the current (drag) or in the opposite direction (anti-drag). This phenomenon  is distinct from the well studied Coulomb drag \cite{nar}, when current in a conductor induces a
current (or voltage) in a nearby conductor. In our case the force is exerted on a small polarizable object, with a well defined excitation, at some frequency $\omega_0$. This can be the resonant frequency of an atom or the frequency of a localized surface plasmon of a nanoparticle. The resulting drag force is a non-monotonic function of the carrier drift velocity $v_0$ and it reaches a maximal value at $v_0$  of the order of $\omega_0 z_0$. The maximal  value of the force is not small, in the sense that it is comparable to the normal (Casimir-Lifshitz) force in equilibrium.

 Formulas for the forces, obtained in the present work, resemble those which appear in the theory of non-contact friction (item (ii) in the Introduction). The two problems, however, are different. In our problem both the particle and the sample are at rest, in the laboratory frame, only the mobile charge carriers are drifting. Our results depend on whether the random spontaneous sources reside predominantly in the lattice or in the electron plasma (Models 1 and 2, respectively). If dissipation in the lattice can be neglected (Model 2) and, moreover, $\epsilon_{L}^{\prime}(\omega)$ is assumed to be constant, then the dielectric function of the medium (lattice + plasma) is a function of $\omega - k_xv_0$ only and, since the random sources are located in the drifting plasma, the situation becomes as close as possible to the case of a medium moving as a whole. However to make the analogy complete one needs an additional strong requirement, namely, that the electrons in the drifting plasma could be considered as being in an internal equilibrium, with some effective temperature $T_{el}$. Otherwise one cannot use Rytov's theory for correlation functions of the random sources.

We limited our considerations to the simplest models and conditions and did not attempt possible generalizations and extensions, like treating the general case (Eq (\ref{epsilons}) with both $\epsilon_{L}^{\prime\prime}$ and $\nu$ finite), or going beyond weak dissipation limit, or including the retardation effects. Finally, let us stress that the high drift velocities, needed to make the discussed effects visible, can be achieved only in materials with low carrier density, like semiconductors, ionic conductors or other types of "bad conductors".

\section{Acknowledgement}
Numerous instructive discussions with J. Avron, J. Feinberg, O. Kenneth and U. Sivan are gratefully acknowledged.
I am indebted to G. Dedkov for sending to me his review, Ref. \cite{ded}, prior to publication.


\begin{thebibliography}{10}
\bibitem{foot1}We use the same letter for a quantity and its Fourier transform but we do indicate explicitely the corresponding
arguments, so there is no confusion.
\bibitem{foot2}$\left\langle \rho_{r}\left(\vec{k}\right)\rho_{r}^{*}\left(\vec{k^{\prime}}\right)\right\rangle
_{\omega}$ here differs from the corresponding expression in
\cite{bs'} by a factor $2\pi$. This is because of different
definitions of the spectral density $\left(x^2\right)_{\omega}$ of
a fluctuating variable $x(t)$. In \cite{bs'}  the spectral
decomposition of the variance $\left\langle x^2(t) \right\rangle$
was taken as $\int d\omega \left(x^2\right)_{\omega}$, while here
we follow the more often used convention $\left\langle x^2(t)
\right\rangle = \int\frac {d\omega}{2\pi}
\left(x^2\right)_{\omega}$. Physical quantities, like various
correlation functions, cannot of course depend on the chosen
definition.
\bibitem{foot3} We take this opportunity to point out that a
factor $(2\pi)^{-2}$ is missing on the LHS of Eq (17) in
\cite{bs'} . This error had no effect on the subsequent
calculations which were based on the correct expression given in
Eq (19) of \cite{bs'}.
\bibitem{foot4}When computing a correlation function for two real
variables, say, $X(t)$ and $Y(t)$, one can of course replace a
variable by its complex conjugate. Therefore in the spectral
decomposition one can have, for instance, $\left\langle
XY^{*}\right\rangle _{\omega}$ or $\left\langle
X^{*}Y\right\rangle _{\omega}$. These quantities differ by the
sign of $\omega$ and, in general, are not equal but after
integrating over  $\omega$ the result is the same.
\bibitem{foot5}We do not consider the possible dependence of the dielectric function and of the particle polarizability on temperature.
\end{thebibliography}
\end{document}